\documentclass[twocolumn,showpacs,aps]{revtex4}

\usepackage{graphicx}%
\usepackage{dcolumn}
\usepackage{amsmath}

\begin{document}

\title{Estimating probabilities from experimental frequencies}

\author{In\'es Samengo}

\email{samengo@cab.cnea.gov.ar}

\affiliation{Centro At\'omico Bariloche and Instituto Balseiro \\
(8400) San Carlos de Bariloche, R\'{\i}o Negro, Argentina}


\begin{abstract}
Estimating the probability distribution ${\bf q}$ governing the
behaviour of a certain variable by sampling its value a finite
number of times most typically involves an error. Successive
measurements allow the construction of a histogram, or frequency
count ${\bf f}$, of each of the possible outcomes. In this work,
the probability that the true distribution be ${\bf q}$, given
that the frequency count ${\bf f}$ was sampled, is studied. Such a
probability may be written as a Gibbs distribution. A
thermodynamic potential, which allows an easy evaluation of the
mean Kullback-Leibler divergence between the true and measured
distribution, is defined. For a large number of samples, the
expectation value of any function of ${\bf q}$ is expanded in
powers of the inverse number of samples. As an example, the
moments, the entropy and the mutual information are analyzed.
\end{abstract}

\maketitle

\section{Estimating probabilities from experimental frequencies}

\label{Intro} The estimation of probability distributions from a
limited number of samples typically involves an error. Consider,
for example, a random variable that can be either 0 or 1, both
values with probability 1/2. An experimenter measures the
variable, say, four times. If $n_0$ (similarly, $n_1$) is the
number of trials the result was 0 (correspondingly, 1), the
possible outcomes are $n_0 = j, n_1 = 4 - j$, where $j$ may vary
between 0 and 4. Each of those possibilities has probability $3 /
2 j! (4 - j)!$ of occurring. If the experimenter estimates the
underlying probability from the frequencies, his or her claim
will be that the probability of getting a zero is $n_0 / 4$.
However, in view that $n_0$ depends on the particular outcome of
the four trials, only a fraction 3/16 of the times will this
procedure give the correct result, that is $f_0 = q_0 = 1/2$.

In the above example, there are three probability distributions
involved. First, there is the {\it true} underlying probability
${\bf q}$, actually governing the outcome of the experiment. In
vector notation, ${\bf q} = (q_0, q_1)$, and in the particular
instance above, ${\bf q} = (1/2, 1/2)$. Then, there is the
frequency count ${\bf f} = (f_0, f_1)$, where $f_i$ is obtained
by dividing $n_i$ by the total number of measurements $N$ (four,
in the example). And finally, there is the probability that ${\bf
f} = {\bf q}$. To define this last probability, one has to
consider all possible samples of $N$ trials, and evaluate how
often the condition ${\bf f} = {\bf q}$ is fulfilled.

More generally, one can define the probability of measuring a
particular ${\bf f}$, while the underlying ${\bf q}$ remains
fixed. This means to consider a probability distribution of all
the possible frequency counts. The independent variable is the
vector ${\bf f}$, which varies in a discrete set, and the
dependent variable is $p({\bf f}|{\bf q})$.

The frequency count ${\bf f}$ is an estimation of the underlying
${\bf q}$. In many applications, however, one is interested not
quite in ${\bf q}$, but rather in some function of ${\bf q}$.
Treves and Panzeri \cite{stefale}, for example, have quantified
the mean error that an experimenter makes when evaluating the
mutual information in the frequency count ${\bf f}$, as an
approximation to that in the true (and unknown) ${\bf q}$. Their
analysis was made in the same spirit as above, that is, they have
considered ${\bf q}$ fixed, while the value of ${\bf f}$ depended
on the particular outcome of $N$ measurements. They have obtained
a clean analytical result, under an independence approximation.
Their approach may be naturally generalized to situations where
${\bf q}$ is a probability density, that is, varies in a
continuous set \cite{network}.

However, what the experimenter knows is not the true ${\bf q}$,
but one particular ${\bf f}$, obtained after $N$ observations.
His or her aim is to estimate the most probable value of ${\bf
q}$ (or of some function of ${\bf q}$) from the knowledge of
${\bf f}$. More generally, the experimenter may be interested in
the whole distribution $P({\bf q} | {\bf f})$, that is, the
probability that the true distribution be ${\bf q}$, given that he
or she has measured ${\bf f}$. This means to settle the problem
the other way round as was studied by Treves and Panzeri, and in
the example above. It actually corresponds to Wolpert and Wolf's
approach \cite{Wolpert} in the estimation of entropies.

In the following section, the properties of the distribution
$P({\bf q} | {\bf f})$ are studied. In Sect. \ref{Tempe}, $P({\bf
q} | {\bf f})$ is written as a Gibbs' distribution, where the
inverse number of samples plays the role of an effective
temperature, and the Kullback-Leibler divergence between ${\bf
f}$ and ${\bf q}$ is the equivalent of the energy of state ${\bf q
}$. As a consequence, a thermodynamic potential is defined, thus
allowing the calculation of the mean Kullback-Leibler divergence
between ${\bf f}$ and ${\bf q}$ by simple derivation. This
inspires the expansion made in Sect. \ref{Expans}, where the
expectation value of an arbitrary function of ${\bf q}$ can be
written as a power series in the inverse number of samples. The
case of the entropy, the mutual information, or any moment of the
distribution ${\bf q}$ is shown in the examples of Sect.
\ref{Ejemplos}. Next, in Sect. \ref{Simul} the analytical results
are confronted with numerical simulations. Finally, in Sect.
\ref{Discu}, the main results are summarized and discussed.

\section{The probability distribution for the true probability distribution}

\label{Formu} Consider the random variable $X$ taking values from
the set ${\bf x} = (x_1, ..., x_S)$, with probabilities ${\bf q} =
(q_1, ..., q_S)$. In principle, there is no need that $x_1, ...,
x_S$ be numerical values, it suffices them to be any exclusive
and exhaustive set of categories.

An experimenter makes $N$ observations of the value of $X$ and
builds a histogram ${\bf n} = (n_1, ..., n_S)$, where $n_i$ is
the number of times the outcome was $x_i$. The experimenter
considers the frequencies ${\bf f} = (f_1, ..., f_S) = (n_1/N,
..., n_S/N)$ as an estimation of the true underlying probability
distribution ${\bf q}$. If the measurements are taken
independently, the probability of measuring ${\bf f}$ given that
the data are sorted according to ${\bf q}$ is equal to the
probability of observing each $x_i$ a number $n_i$ of times, that
is,
\begin{equation}
p({\bf f} | {\bf q}) = N! \ \Pi_i \frac{q_i^{n_i}}{n_i!} =
\frac{N!} { \Pi_i (N \, f_i)!} \exp\left(N \sum_i f_i \ln q_i
\right) . \label{multinomial}
\end{equation}
However, the knowledge the experimenter has at hand is ${\bf f}$,
not ${\bf q}$. He or she may therefore wonder what is the
probability that the true distribution be ${\bf q}$, given that
the outcome of the experiment was ${\bf f}$. This means to
evaluate a probability density $P({\bf q}|{\bf f})$, whose
independent variable ${\bf q}$ runs over all the possible
distributions of the data. That is, all vectors in $\Re ^S$ such
that
\begin{eqnarray}
\sum_i q_i &=& 1 \nonumber \\
0 \le &q_i& \le 1, \ \ \ \forall i. \label{dominio}
\end{eqnarray}
The set of all ${\bf q}$ obeying Eqs. (\ref{dominio}) constitutes
the domain ${\cal D}$ where $P({\bf q} | {\bf f})$ is defined. It
is a finite portion of an $(S - 1)$-dimensional plane embedded in
$\Re ^S$, and is normal to the vector $(1, 1, ..., 1)$.

Notice that since each $f_i$ is the ratio of two natural numbers,
the set of possible frequencies ${\bf f}$ is discrete. The domain
${\cal D}$, on the contrary, contains a continuum of distributions
${\bf q}$. Consequently, $p({\bf f} | {\bf q})$ is a probability,
whereas $P({\bf q} | {\bf f})$ is a density.

Bayes' rule states that
\begin{equation}
P({\bf q} | {\bf f}) = \frac{p({\bf f} | {\bf q}) P({\bf q})}
{p({\bf f})},\label{bayes}
\end{equation}
where $P({\bf q})$ is the prior probability distribution for
${\bf q}$, and
\begin{equation}\label{total}
p({\bf f}) = \int_{{\cal D}}  P({\bf f}|{\bf q}) \ P({\bf q}) \
dS_{\bf q} \ .
\end{equation}
Here, $dS_{{\bf q}}$ is a volume element, in ${\cal D}$.

The prior $P({\bf q})$ contains all additional pieces of
knowledge about ${\bf q}$, apart from the experimental data.
Here, the assumption is made that there is {\it no} a priori
knowledge. However, it turns out to be crucial to specify {\it
what} is it that is not known \cite{Ilya}. A prior that is uniform
over ${\cal D}$, as was used by Wolpert and Wolf \cite{Wolpert},
is certainly not uniform over any non linear function of ${\bf
q}$, for example the log-likelihood. Thus, not knowing anything
about ${\bf q}$ implies knowing something about $\ln q$, which in
turn may result in awkward scaling properties. In this work, the
power prior
\begin{equation}
P_\beta({\bf q}) = \frac{\Pi_{i = 1}^S \ q_i^{\beta - 1}}{{\cal
Z}_\beta}, \label{prior}
\end{equation}
is repeatedly used, with ${\cal Z}_\beta = \sqrt{S}
[\Gamma(\beta)]^S / \Gamma(S\beta)$ (notice that when $\beta \to
0, {\cal Z}_\beta \to \sqrt{S}$). However, as was shown in
\cite{Ilya} choosing any of these priors results in a surprisingly
peaked a priori distribution of the possible entropies. Hence,
the choice of the prior is a delicate issue and, in any
particular application, it should be done carefully. Here, no
attempt will be made to instruct on the way such a choice should
be made, but since the results that follow are strongly grounded
on Bayesian inference, their validity is, at most, ´´as good as
the prior'' \cite{Wolpert}.

Replacing Eqs. (\ref{multinomial}) and (\ref{total}) in Eq.
(\ref{bayes}),
\begin{equation}
P({\bf q} | {\bf f}) = \frac{\exp\left[-N D({\bf f}, {\bf
q})\right] P({\bf q})}{{\cal Z}}, \label{prob1}
\end{equation}
where $D$ is the Kullback-Leibler divergence between ${\bf f}$ and
${\bf q}$
\begin{equation}
D({\bf f}, {\bf q}) = \sum_{i} f_i \ln \left(\frac{f_i}{q_i}
\right),
\end{equation}
and quantifies is the mean information for discriminating in
favor of ${\bf f}$ against ${\bf q}$, given the data \cite{Kull}.
The function ${\cal Z}$ reads
\begin{equation}
{\cal Z} = \int_{{\cal D}} dS_{\bf q} \ P({\bf q}) \ \exp \left[-N
D({\bf f}, {\bf q})\right]. \label{particion}
\end{equation}

In the remaining of the section, the properties of $P({\bf q}|{\bf
f})$ are studied for the particular $P_\beta({\bf q})$ defined in
Eq. (\ref{prior}). In doing so, the integral
\begin{equation}
\int_{{\cal D}} \Pi_{i = 1}^S \  q_i^{m_i} \ dS_{{\bf q}} =
\sqrt{S} \ \frac{\Pi_i \ \Gamma(m_i + 1)}{\Gamma(S + \sum_i m_i)},
\label{integrales}
\end{equation}
is frequently encountered. Equation (\ref{integrales}) was first
derived in \cite{Wolpert}, and an alternative proof may be found
in the Appendix.

For the priors in Eq. (\ref{prior}), the function ${\cal Z}$ Eq.
(\ref{particion}) may be calculated analytically, and it reads
\begin{equation}
{\cal Z} = \exp \left[ N {\cal H}({\bf f}) \right]  \ \sqrt{S} \
\frac{\Pi_{j = 1}^S \ \Gamma(N f_k + \beta)}{\Gamma(N + S
\beta)}, \label{parti}
\end{equation}
where ${\cal H}$ is the entropy of a distribution
\begin{equation}
{\cal H}({\bf f}) = -\sum_{i = 1}^S f_i \ln f_i. \label{entropia}
\end{equation}
Thus, replacing Eq. (\ref{parti}) in Eq. (\ref{prob1})
\begin{equation}
P({\bf q} | {\bf f}) = \frac  {\Gamma(N + S \beta)} {\sqrt{S}}
\Pi_i \frac{q_i^{N f_i + \beta - 1}}{\Gamma(N f_i + \beta)}.
\label{result}
\end{equation}

The most probable ${\bf q}^M = (q_1^M, ..., q_S^M)$ is obtained by
maximizing Eq. (\ref{result}), under the normalization constrain.
The result is
\begin{equation}
q_i^M = \frac{N f_i + \beta - 1}{N + S(\beta - 1)}. \label{max}
\end{equation}
Thus, if $P({\bf q})$ is uniform in ${\cal D}$ ($\beta = 1$),
then the most probable ${\bf q}$ is ${\bf f}$. With the maximum
likelihood prior ($\beta \to 0$), the most probable ${\bf q}$ is
shifted from ${\bf f}$ towards lower counts. The
Krichevsky-Trofimov estimator \cite{Kirch}($\beta = 1/2$) and the
Shurmann-Grassberger \cite{Shur} $\beta = 1 / S$ lie in between.

\begin{figure}[htbf]
\begin{center}
\resizebox{\columnwidth}{!}{\includegraphics{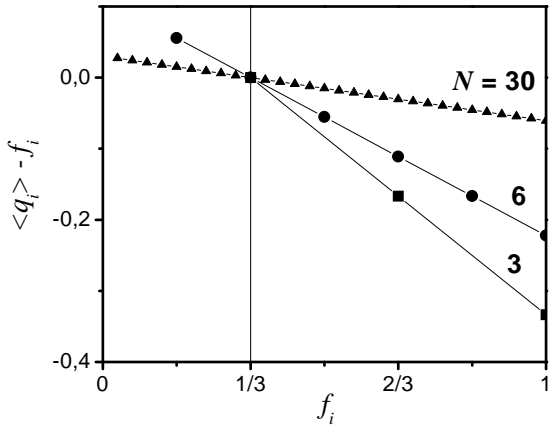}}
\end{center} \caption{Difference between $\langle q_i
\rangle$ and $f_i$, as a function of $f_i$. The value of $\beta$
has been set to 1. The three lines correspond to $N$ = 3, 6 and
30. Here, $X$ may take 3 values ($S$ = 3). When $f_i < 1/3$, the
expectation value of $q_i$ is larger than the measured frequency
$f_i$. As $N$ increases, the effect becomes less important.}
\label{fig1}
\end{figure}

Using Eq. (\ref{integrales}) the expectation value of each
component $q_i$ may be calculated,
\begin{equation}
\langle q_i \rangle = \frac{N f_i + \beta}{N + S \beta}.
\label{cada}
\end{equation}
For the uniform prior $\beta = 1$, this equation reduces to
Laplace's estimator of probabilities, first introduced by in his
{\it Essay on probabilities}. In figure \ref{fig1}
the difference between $\langle q_i \rangle$ and the frequency
count $f_i$ is shown, for $\beta = 1$. It is seen that when $f_i$
is smaller than $1 / S$, $\langle q_i\rangle$ is larger than
$f_i$. On the other hand, if $f_i > 1 / S$, then $\langle q_i
\rangle < f_i$. That is, the mean value of $q_i$ is displaced
from the frequency count so as to approach the flat distribution
$1 / S$. Of course, the larger the number of samples $N$, the
smaller the effect. Changing the value of $\beta$ is equivalent
to re-scaling the vertical axis of figure \ref{fig1}.

Typically, one wants to make a guess about the true ${\bf q}$.
Here, two possible estimators have been calculated: the maximum
${\bf q}^M$ and the mean $\langle {\bf q} \rangle$. By using the
maximum, one is choosing the value that is most probably correct.
But of course, eventually one will also make an error. If one
measures the error as a $({\bf q}^M - {\bf q})^2$, and averages it
with $P({\bf q} | {\bf f})$, its mean turns out to be larger than
if one had chosen $\langle {\bf q} \rangle$ \cite{Wolpert}. Hence,
although ${\bf q}^M$ is the estimator that gives the correct
answer most frequently, if one cares for the typical size of the
errors, $\langle {\bf q} \rangle$ is a better choice.

When using $\langle {\bf q} \rangle$ as an estimator, the
covariance matrix $\Sigma_{ij}$ may be of interest. By means of
Eq. (\ref{integrales})it is easy to show that for $i \ne j$
\begin{eqnarray}
\Sigma_{ij} &=& \left \langle (q_i - \langle q_i \rangle)(q_j -
\langle q_j \rangle) \right \rangle  \label{otra} \\
&=& - \frac {(N f_i + \beta)(N f_j + \beta)} {(N + S\beta)^2(N +
S\beta + 1)} \nonumber \\ & & \rightarrow - \frac {f_i f_j} {N} \
{\rm when \ } N \gg S, \nonumber
\end{eqnarray}
whereas for $i = j$
\begin{eqnarray}
\Sigma_{ii} &=& \left \langle (q_i - \langle q_i \rangle)^2 \right
\rangle = \label{una} \\ & & \frac {(N f_i + \beta)[N(1 - f_i) +
\beta(S - 1)]} {(N + S \beta)^2(N + S\beta + 1)} \nonumber \\ & &
\rightarrow \frac {f_i(1 - f_i)} {N} \ {\rm when \ } N \gg S,
\nonumber
\end{eqnarray}
The negative sign in Eq. (\ref{otra}) derives from the
normalization condition: since the sum of all $q_i$ is fixed to
unity, if one of them surpasses its mean, it is to be expected
that some other component will be below. In contrast, Eq.
(\ref{una}) shows that $\Sigma_{ii}$ is always positive.

The expectation value of ${\bf q}$ Eq. (\ref{cada}) together with
the covariance matrix Eqs. (\ref{otra}) and (\ref{una}) are
useful to give the Gaussian approximation to $P({\bf q} | {\bf
f})$, centered in its mean:
\begin{equation}
P({\bf q} | {\bf f}) = K \  \exp \left[ - \frac{1}{2} ({\bf q} -
\langle{\bf q} \rangle)^t \ \tilde{\Sigma}^{-1}({\bf q} -
\langle{\bf q} \rangle)\right], \label{guau}
\end{equation}
where the super-script $t$ means transposed, and $K$ is a
normalization constant. Equation (\ref{guau}) is only defined in
the plane containing ${\cal D}$, normal to the vector $(1, 1,
..., 1)$. Actually, $\Sigma$ does not have an inverse in the
entire space $\Re^S$, since the direction $(1, 1, ..., 1)$ is one
of its eigenvectors,  with eigenvalue equal to zero. However,
being $\Sigma$ a symmetric matrix, it can be diagonalized by an
orthogonal basis. Hence, the $S - 1$ remaining eigenvectors lie in
the plane containing ${\cal D}$. The restriction of $\Sigma$ into
that subspace is $\tilde{\Sigma}$, and its inverse is the matrix
appearing in the exponent of Eq. (\ref{guau}).

In order to normalize the approximation (\ref{guau}) an integral
of a Gaussian function in ${\cal D}$ is needed. This is certainly
not an easy task. If, however, one can assume that the
distribution is sufficiently peaked so that $P({\bf q} | {\bf f})
\approx 0$, for ${\bf q}$ in the border of ${\cal D}$, then the
domain ${\cal D}$ can be extended to the whole plane normal to
$(1, 1, ..., 1)$. In that case, $K^{-1} = \sqrt{2 \pi \Pi_j
\lambda_j}$, where $\lambda_j$ are the $S - 1$ eigenvalues of
$\tilde{\Sigma}$. While the calculation of all the $\lambda_j$ is
a difficult problem, it is quite straightforward to show that
when $N \gg S$, all the $\lambda_j$ are proportional to $1 / N$.
Therefore, the square root of each eigenvalue is a useful measure
of the width of $P({\bf q} | {\bf f})$ in the direction of its
eigenvector.

However, the Gaussian approximation (\ref{guau}) is not useful for
other purposes, as for instance, calculating mean values, since
it lacks from analytical expressions as (\ref{integrales}). As a
consequence, in what follows, the full Eq. (\ref{result}) is used.

Equation (\ref{integrales}) allows the evaluation of all moments
of $P(q_i|{\bf f})$
\begin{equation}
\langle q_i^k \rangle = \frac{\Gamma(N f_i + k + \beta) \Gamma(N +
S\beta)}{\Gamma(N f_i + \beta) \Gamma(N + S\beta + k)}.
\label{medio}
\end{equation}
Since the moments are the coefficients of the Taylor expansion of
the Fourier transform of a distribution, the single-component
distribution reads
\begin{eqnarray}
P(q_i | {\bf f}) &=& P(q_i | f_i) \label{completo} \\
&=& \frac{q^{N f_i + \beta - 1} (1 - q)^{N(1 - f_i) + \beta(S - 1)
- 1}}{B[N f_i + \beta, N(1 - f_i) + \beta(S - 1)]}, \nonumber
\end{eqnarray}
where $B(x, y) = \Gamma(x)\Gamma(y) / \Gamma(x + y)$. Figure
\ref{fig2} displays the distribution $P(q_i | f_i)$ for three
different values of $N$, and $\beta = 1$. In all cases, when $N$
is large, the distribution is symmetrical, and reaches its
maximum value in $q_i = f_i = 1/3$. In fact, it may be shown
analytically that when $N \gg 1$,
\begin{equation}
\lim_{N \gg 1} P(q_i | f_i) = \frac{1}{\sqrt{2 \pi \sigma^2}} \
\exp \left[ -(q_i - f_i)^2 / 2 \sigma^2 \right],
\end{equation}
where $\sigma = [f_i (1 - f_i)/ N]^{1/2}$. That is, the
distribution tends to a Gaussian function centered at the
experimental frequency, and with a mean dispersion that
diminishes with the square root of the number of samples. Notice
that in this limit, $P({\bf q}|{\bf f})$ does not depend on
$\beta$.

It may be seen in Fig. \ref{fig2} that for smaller values of $N$,
the distribution is no longer symmetrical. In fact, since $S = 2$
and $f_1 = 1/3 < 1/S$, the tail in $P(q_1 | f_1)$ extends to the
right, resulting in a positive $\langle q_i \rangle - f_i$, as
predicted by equation (\ref{medio}).

\section{The inverse number of samples as an effective temperature}

\label{Tempe}

Equation (\ref{prob1}) states that $P({\bf q} | {\bf f})$ is
completely analogous to a Gibbs distribution, where the number of
samples $N$ plays the role of the inverse of the temperature,
$D({\bf f}, {\bf q})$ is the equivalent to the energy of the
state ${\bf q}$, and $P({\bf q})$ is the density of states. This
analogy was first pointed out in the context of machine learning
\cite{Seung}, and since then, several times in learning theory
(see for example \cite{Bill}). In these cases, when fluctuation
where neglected, the probability distribution under study had the
form of Eq. (\ref{prob1}). In the present context, no
approximations are needed to write down Eq. (\ref{prob1}).

The exponential factor in (\ref{prob1}) depends on ${\bf q}$ and
${\bf f}$ only in the combination $D({\bf f}, {\bf q})$,
diminishing exponentially as the divergence between the two
distributions grows. Its maximum is attained when $D = 0$. It can
be shown \cite{Kull} that for any ${\bf f}$ and ${\bf q}$,
$D({\bf f}, {\bf q}) \ge 0$, and the equality holds only when
${\bf f} = {\bf q}$.

Defining the thermodynamic potential
\begin{equation}
F = - \ln {\cal Z}
\end{equation}
it follows that
\begin{eqnarray}
\langle D \rangle &=& \frac{\partial F}{\partial N}, \\
\sigma_D^2 &=& \left\langle D^2 - \langle D \rangle ^2 \right
\rangle = - \frac {\partial^2 F}{\partial N^2},
\end{eqnarray}
where the mean values $\langle (\cdot) \rangle$ are defined by
$\int_{{\cal D}} (\cdot) P({\bf q} | {\bf f}) dS_{{\bf q}}$.

For example, when the prior is given by Eq. (\ref{prior}),
\begin{equation}
\langle D \rangle = {\cal H}({\bf f}) - \Psi(N + S\beta) + \sum_i
f_i \Psi(N f_i + \beta), \label{dmed}
\end{equation}
where $\Psi(x) = {\rm d}\ln \Gamma(x)/{\rm d}x$ is the Digamma
function \cite{abra}. It is easy to show that
\begin{equation}
\lim_{N \gg S} \langle D \rangle = \frac{S - 1}{2 N} + {\cal
O}(1/N^2). \label{med}
\end{equation}
Here, both $N$ and $N f_i$ have been supposed large, for all $i$.
Since $f_i$ is of the order of $1/S$, the above limit holds when
$N \gg S$. Equation (\ref{med}) states that for a large number of
samples, the expected value of the divergence between the
experimental frequencies and the true distribution does not
depend on the measured ${\bf f}$. It grows linearly with the
number of items, and decreases as $1 / N$.

Accordingly,
\begin{equation}
\sigma^2_D = - \Psi^1(N + S\beta) + \sum_{i = 1}^S f_i^2 \Psi^1(N
f_i + \beta),
\end{equation}
where $\Psi^1(x) = {\rm d}\Psi(x)/{\rm d}x$, is the first
Polygamma Function \cite{abra}. Taking the limit of a large
number of samples,
\begin{equation}
\lim_{N \gg S} \sigma^2_D = \frac{S - 1}{2 N^2} + {\cal O}(1/N^3).
\end{equation}
In the limit $N \gg S$, the mean quadratic dispersion does not
depend on the measured $f_i$.

\section{Estimation of functionals of ${\bf q}$, for a large
number of samples.}

\label{Expans}

Many times, one is interested in the value of some function
$W({\bf q})$. For instance, if $X$ takes numerical values, $W$
may be the mean $\bar{X} = \sum_i x_i q_i$. Or, in some other
application, $W$ may be the entropy of the distribution ${\bf q}$
(see equation (\ref{entropia})).  If the set $X$ is the Cartesian
product of two other sets $X = Z^1 \times Z^2$, such that
$\forall x_i \in X: x_i = (z^1_a, z^2_b)$, where $z^1_a \in Z^1$
and $z^2_b\in Z^2$, then $W$ may be the mutual information $I$
between $Z^1$ and $Z^2$:
\begin{equation}
I = \sum_{a b} q_{a b} \ln \left[ \frac {q_{a b}} {q_{a.} q_{.b}}
\right], \label{imutua}
\end{equation}
where
\begin{eqnarray}
q_{a.} &=& \sum_{b} q_{ab},
\nonumber \\
q_{.b} &=& \sum_{a} q_{ab} \ \ . \label{archi}
\end{eqnarray}

Since ${\bf q}$ is unknown, an interesting guess for $W({\bf q })$
is its Bayesian estimation
\begin{equation}
\langle W \rangle = \int_{\cal D} W({\bf q}) P({\bf q} | {\bf f}),
\label{Wmed}
\end{equation}
which has the appealing property of minimizing the mean square
error \cite{Wolpert}. The zero order guess for $\langle W \rangle$
is $W({\bf f})$. In what follows, a systematic method to improve
this value is derived.

In the previous section the expectation value of the divergence
between the true and the measured distribution was calculated, as
well as the size of the fluctuations, for the priors in Eq.
(\ref{prior}). As the number of samples increases, both the
expected divergence and the fluctuations diminish as $1/N$. Since
a small divergence means that the two distributions are
necessarily very similar, only the ${\bf q}$ that are very near
${\bf f}$ have a non vanishing probability---for $D$ sufficiently
small, this argument holds for any definition of similarity.

As a consequence, it is reasonable to expand $W({\bf q})$ in its
Taylor series in the neighborhood of ${\bf f}$. Hence, Eq.
(\ref{Wmed}) reads
\begin{equation}
\langle W \rangle = \left \langle \sum_{k = 0}^\infty \frac {1}
{k!} \left( \sum_{i = 1}^S (q_i - f_i) \frac {\partial} {\partial
q_i} \right)^k \left. W \right|_{{\bf f}} \right \rangle.
\label{taylor}
\end{equation}
Since $P({\bf q} | {\bf f})$ decreases dramatically as ${\bf q}$
departs from ${\bf f}$, the higher order terms (large $k$) in Eq.
(\ref{taylor}) should become negligible, at least, for large $N$.

In the first place, the mean values of Eq. (\ref{taylor}) are
evaluated for the special case of the power law priors. This
involves, basically, the computation of integrals in ${\cal D}$
of $\Pi_{i = 1}^S (q_i - f_i)^{k_i}$, for a set of non negative
indexes $(k_i, k_2, ... k_S)$ that sum up to $K$. This can be
done using Eq. (\ref{integrales}). Of course, the term $k =
0$---that is, the raw guess--does not depend on $N$. It may be
shown that only $k = 1$ and $k = 2$ are proportional to $1/N$.
Specifically,
\begin{eqnarray}
\left \langle  q_i - f_i \right \rangle &=& \frac {\beta(1 - S
f_i)} {N + S\beta} \nonumber \\ & & \rightarrow \frac {\beta(1 - S
f_i)} {N}, \ \ {\rm when \ } N \gg S.
\end{eqnarray}
In the same way , if $i \ne j$
\begin{eqnarray}
& & \left \langle (q_i - f_i)(q_j - f_j) \right \rangle =
\label{corrij} \\ & &  - \frac {N f_i f_j - \beta \left[ \beta +
(1 + S \beta) (S f_i f_j - f_i - f_j) \right]} {(N + S\beta) (N +
S\beta + 1)} \nonumber \\ & & \rightarrow - \frac{f_i f_j}{N} \ \
\ {\rm when \ } N \gg S, \nonumber
\end{eqnarray}
whereas when $i = j$
\begin{eqnarray}
& & \left \langle (q_i - f_i)^2 \right \rangle = \label{autoc} \\
& & \frac {N f_i (1 - f_i) + \beta[1 + \beta + f_i(1 + S \beta)(S
f_i - 2)]} {(N + S \beta) (N + S \beta + 1)} \nonumber \\ & &
\rightarrow \frac{f_i(1 - f_i)}{N} \ \ \ {\rm when \ } N \gg S.
\end{eqnarray}

Summarizing, to first order in $1/N$,
\begin{eqnarray}
\left \langle W \right \rangle &\approx& W\left({\bf f}\right) +
\label{gener} \\ & & +
 \sum_{i = 1}^S \left. \frac {\partial W}
{\partial q_i} \right|_{{\bf f}} \frac {\beta(1 - S f_i)} {N} +
\nonumber \\ & & + \frac{1}{2} \sum_{i = 1}^S \left. \frac
{\partial^2 W} {\partial
q_i^2} \right|_{{\bf f}} \frac {f_i (1 - f_i)} {N} - \nonumber \\
& & - \sum_{i = 1}^S \sum_{j < i} \left. \frac {\partial^2 W}
{\partial q_i \partial q_j} \right|_{{\bf f}} \frac {f_i f_j}
{N}. \nonumber
\end{eqnarray}
This general formula allows the calculation of the first
correction of the expectation value of an arbitrary function
$W({\bf q})$, whenever the prior is given by Eq. (\ref{prior}).

Now, consider the more general case of an arbitrary prior. If
$P({\bf q})$ is not given by Eq. (\ref{prior}), then one can
still proceed as above, but replacing $W({\bf q})$ by the product
$W({\bf q})P({\bf q})$, and setting $\beta = 1$.

\section{Examples}

\label{Ejemplos} Here, the expansion (\ref{gener}) is applied to
a few particular cases. Wolpert and Wolf \cite{Wolpert} have
already calculated the first two examples exactly (Subsect.
\ref{ejuno} and \ref{ejdos}), in the particular case of $\beta =
1$. Their results, once expanded up to first order in $1/N$ are
now compared to Eq. (\ref{gener}), for verification. The
advantage of Eq. (\ref{gener}) is that, in contrast to Wolpert and
Wolf's approach, it applies to any function $W$. The counterpart,
of course, is that it gives no more than the first correction to
$\langle W \rangle$. Subsection \ref{ejtres} deals with the
calculation of moments.

\subsection{The mean value of the entropy}

\label{ejuno}

In the first place, the function $W({\bf q})$ is taken to be the
entropy ${\cal H}$ of the distribution ${\bf q}$, defined in Eq.
(\ref{entropia}), for ${\bf q} = {\bf f}$. It is easy to verify
that $\partial {\cal H} / \partial q_i = - [1 + \ln q_i]$,
whereas $\partial^2 {\cal H} / \partial q_i \partial q_j =
-\delta_{ij} / q_i$, where $\delta_{ij}$ is Kroeneker delta
function: $\delta_{ij} = 1$, if $i = j$ and $\delta_{ij} = 0$, if
$i \ne j$. Replacing in Eq. (\ref{gener}) and keeping only up to
the first order in $1/N$ one arrives at
\begin{eqnarray}
\left \langle {\cal H} \right \rangle &=& \left(1 - \frac {\beta
S} {N} \right) {\cal H}({\bf f}) + \label{juli} \\
& & \frac {\beta} {N} \sum_{i = 1}^S \ \ln \left( \frac{1}{f_i}
\right) - \frac{S - 1}{2 N} + {\cal O}(1/N^2).
\end{eqnarray}
For the case of $\beta = 1$, this same expression is obtained by
expanding the exact result, obtained in \cite{Wolpert}
\begin{eqnarray}
\left\langle {\cal H} \right \rangle_{[3]} &=& - \sum_{i = 1}^S
\frac {N f_i + 1} {N + S} \left[ \Phi^{(1)}(N f_i + 2) - \right.
\nonumber \\
& & \left. \Phi^{(1)}(N + S + 1) \right],
\end{eqnarray}
where $\Phi^{(1)}(x) = {\rm d}\ln \Gamma(x) / {\rm d} x$ is the
Digamma function \cite{abra}.

\subsection{The mean value of the mutual information}

\label{ejdos}

Now $W$ is taken to be the mutual information between two sets, as
defined by Eq. (\ref{imutua}). Replacing in Eq. (\ref{gener}),
\begin{eqnarray}
\left\langle I \right\rangle &=& I({\bf f}) \left(1 - \beta
\frac{S_1 S_2}{N} \right) + \label{desi} \\
& &\frac {S_1 S_2 + 1 - S_1 - S_2} {2 N} + \frac{\beta}{N} \sum_{a
b} \ln \left( \frac {f_{a b}} {f_{a.} f_{.b}} \right), \nonumber
\end{eqnarray}
Where $S_1$ and $S_2$ are the number of elements in the sets $Z^1$
and $Z^2$. When $\beta = 1$, Eq. (\ref{desi}) coincides with the
expansion up to first order in $1/N$ of the exact result derived
in \cite{Wolpert},
\begin{eqnarray}
\langle I \rangle_{[3]} &=& \sum_{a b} \frac {N f_{a b} + 1} {N +
S_1 S_2} \left[ \Phi^{(1)}(N f_{a b} + 2) - \right. \nonumber
\\ & &  \left. \Phi^{(1)}(N + S_1 S_2 + 1) \right] -
\nonumber \\
&-&  \sum_{a} \frac {N f_{a.} + S_2} {N + S_1 S_2} \left[
\Phi^{(1)}(N f_{a.} + S_2 + 1)\right. - \nonumber \\
& & \left.\Phi^{(1)}(N + S_1 S_2 + 1) \right] - \nonumber \\
&-&  \sum_{b} \frac {N f_{. b} + S_1} {N + S_1 S_2} \left[
\Phi^{(1)}(N f_{. b} + S_1 + 1)\right. - \nonumber \\
& & \left. \Phi^{(1)}(N + S_1 S_2 + 1) \right]. \label{lio}
\end{eqnarray}
The quantities $f_{a.}$ and $f_{.b}$ in Eqs. (\ref{desi}) and
(\ref{lio}) are defined as in (\ref{archi}).

In contrast to the result obtained in \cite{stefale}, the first
order correction to the mutual information does bear a dependence
on the values of the individual probabilities $f_{a b}$. There is
no conflict, however, between the two results, since the mean
value in Eq. (\ref{desi}) involves the distribution $P({\bf
q}|{\bf f})$. The approach in \cite{stefale}, instead, uses
$p({\bf f} | {\bf q})$, while the true ${\bf q}$ is fixed. In the
present approach, the mean value $\langle I \rangle$ can be
either higher or lower than $I({\bf f})$.

\subsection{The mean value of functions of $X$}

\label{ejtres}

Consider a function $g:\{x_1, ..., x_S \} \rightarrow {\cal R}$
that maps the possible values of $X$ into real numbers. For
example, if $X$ takes numerical values, then $g_k$ can be such
that $g_k(x_i) = x_i^k$. For each such $g$, another function
$G:{\cal D} \rightarrow {\cal R}$ is defined, namely $G({\bf q})
= \sum_i g(x_i) q_i$. In the example above, $G_k$ is the
$k$-moment of the distribution ${\bf q}$. The expectation value
$\langle G \rangle$ is easily calculated using Eq. (\ref{gener}),
and reads
\begin{equation}
\langle G \rangle = G({\bf f}) \left(1 - \frac{\beta S}{N} \right)
+ \frac{\beta}{N} \sum_{i= 1}^S g(x_i).
\end{equation}
In particular, for the $g_k$ considered above, this is the first
order correction to all moments of ${\bf q}$.

\section{Numerical simulations}

\label{Simul} In this section, Eq. (\ref{gener}) is confronted to
the result of numerical simulations. Once again, and just to
follow previous studies, $W({\bf q})$ is set equal to the mutual
information. However, in contrast to what was done up to now
\cite{Wolpert,stefale,network}, the simulations are performed
strictly within the present framework. That is, the measured
frequency ${\bf f}$ is kept fixed, and the probability for the
true ${\bf q}$ is evaluated.

The procedure to measure numerically $P({\bf q} | {\bf f})$ is now
explained. As before, $X$ takes values in a set of $S$ elements.
Hence, ${\bf f}$ and ${\bf q}$ are $S$-dimensional vectors. The
value of ${\bf f}$ is fixed. The domain ${\cal D}$ is discretized
into a number $J$ of cells. Each cell corresponds to a vector
${\bf q}$ that will be visited by the program. The larger the
number of cells $J$, the better the sampling of the domain ${\cal
D}$. For each one of these cells, the value of $X$ is measured
$N$ times. The outcomes are sorted with the distribution ${\bf
q}$ of the actual cell. If the frequency count thus obtained
equals ${\bf f}$, the counter of the selected cell is increased
(there is counter for each cell in ${\cal D}$). The comparison
between the frequency count and the (fixed) ${\bf f}$ is done
with precision $\epsilon$. The procedure is repeated $M$ times
($M$ large) in order to have enough counts. This algorithm allows
to construct a histogram for the probability that a given ${\bf q}
\in {\cal D}$ generates the selected ${\bf f}$.

For simplicity, in the results below the number of trials $M$ is
the same for all cells. This is equivalent to using a uniform
prior in ${\cal D}$ ($\beta = 1$). A simulation with a non uniform
prior can be carried out by choosing a different $M$ for each
cell.

The two parameters that determine the precision of the
simulations are $J$ and $\epsilon$. If $D_J$ is the
Kullback-Leibler divergence between two neighboring ${\bf q}$
cells, whenever $1 / N \ll D_J$ then the only vector ${\bf q}$
that produces frequency counts equal to ${\bf f}$ is ${\bf q} =
{\bf f}$. That is, for $N$ sufficiently large, the discretized
system behaves as if $N = \infty$. Notice that for large $J$, two
neighboring cells correspond to ${\bf q}$ and ${\bf q} +
\delta{\bf q}$, with each $\delta q_i \propto J^{S - 1}$. Thus,
the Kullback-Leibzig distance between the two is $\approx S/J^{S
- 1}$. This means that when $N$ reaches $J^{S - 1}/S$, the
simulation starts to behave as if $N$ were actually infinite.

On the other hand, if $\epsilon$ is not small enough, one
mistakenly counts coincidences with ${\bf f}$, just because the
criterion used in the comparison is too brute. In other words, a
large $\epsilon$ allows that cells ${\bf q}$ too far away from
${\bf f}$ do give rise to frequency counts equal to ${\bf f}$.
That is, the system behaves as if $N$ where smaller than its
actual value.

\begin{figure}[htbf]
\begin{center}
\resizebox{\columnwidth}{!}{\includegraphics{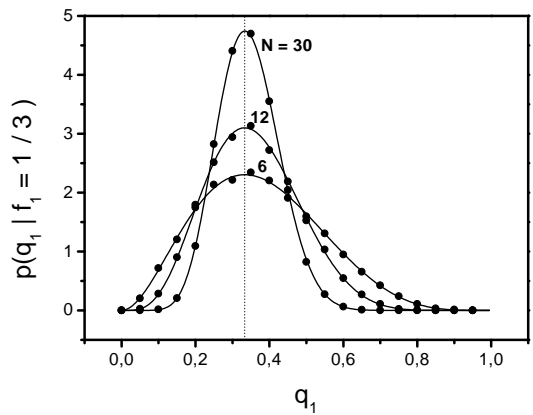}}
\end{center}
\caption{Probability distribution $P(q_1 | f_1)$ for the case
$f_1 = 1/3$, $\beta = 1$ and $S = 2$. Different curves correspond
to several values of the number of samples $N$. The full line
depicts the analytical result Eq. (\ref{completo}), while the
dots are the numerical simulations (see Sect. \ref{Simul}).}
\label{fig2}
\end{figure}

The dots in figure \ref{fig2}
show the result of the above procedure, for a single component
$q_1$. As observed, there is very good agreement with the full
line, showing the analytical result, Eq. (\ref{result}).

\begin{figure}[htbf]
\begin{center}
\resizebox{\columnwidth}{!}{\includegraphics{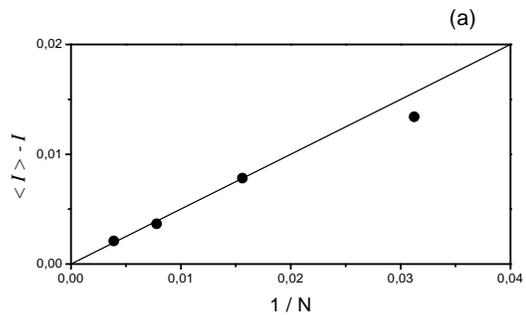}}
\end{center}
\begin{center}
\resizebox{\columnwidth}{!}{\includegraphics{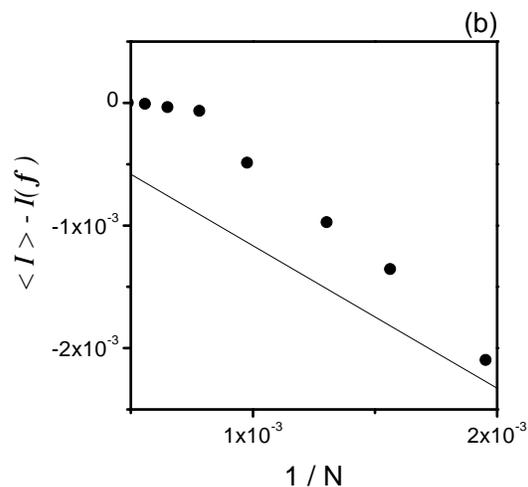}}
\end{center} \caption{Difference between the expectation value of the mutual
information $\langle I \rangle$ and the measured $I({\bf f})$, as
a function of the inverse number of samples $1/N$. The $\beta =
1$ prior was considered. The full line represents the analytical
result, Eq. (\ref{desi}), and the dots the simulations. In (a),
$f_{11} = f_{12} = f_{21} = f_{22} = 1/4$, and $I({\bf f}) = 0$.
For each cell in ${\cal D}$, 30,000 sets of $N$ samples have been
sorted. In (b), $f_{11} = 0.4$, $f_{12} = 0.1$, $f_{21} = 0.1$,
and $f_{22} = 0.4$, so $I({\bf f}) = 0.192745$. For each cell in
${\cal D}$, 10,000 sets of $N$ samples have been sorted. In both
cases, each axis in ${\bf q}$ space has been divided in 20
intervals, in order to discretize ${\cal D}$, while the parameter
$\epsilon$ was set to 0.0125.} \label{fig3}
\end{figure}

To evaluate the expectation value of a certain function, one
simply needs to calculate the sum
\begin{equation}
\left. \langle W \rangle \right|_{\rm numerical} = \sum_{{\rm
cells \ in \ {\cal D}}} W({\bf q}) P({\bf q} | {\bf f}),
\label{numer}
\end{equation}
using the $P({\bf q} | {\bf f})$ obtained with the algorithm
explained above. Figure \ref{fig3} depicts the result for the
mutual information,  with $\beta = 1$. The dots represent the
simulations, Eq. (\ref{numer}), whereas the full line shows the
analytical result (\ref{desi}). The computational time required
to evaluate $P({\bf q} | {\bf f})$ increases exponentially with
the number of  dimensions $S$. Hence, in the present comparison
it is desirable to keep $S$ as small as possible. However, in
order to define a mutual information two sets $Z^1$ and $Z^2$ are
needed, with $S_1$ and $S_2$ elements each. In figure \ref{fig3},
$S_1 = 2$ and $S_2 = 2$, thus making a 3 dimensional domain
${\cal D}$.

In (a) the selected ${\bf f}$ had no mutual information: $I({\bf
f}) = 0$. The graph shows that the expectation value of $I$ is
positive. With the chosen parameters (see the caption of the
figure), the analytical result (\ref{desi}) coincides exactly
with the one derived by Treves and Panzeri \cite{stefale}, that
is, $\langle I \rangle = (S_1 - 1)(S_2 - 1) / 2N$. Since for
$I({\bf f}) = 0$, Eq. (\ref{desi}) reduces to $\langle I \rangle =
S_1 S_2 + 1 - S_1 - S_2 / 2N$, for some particular choices of
$S_I$ and $S_J$ the two expressions may coincide. It should be
kept in mind, however, that this is just a coincidence, and the
two mean values have different meanings.

In contrast, in case (b) the value of $I({\bf f})$ is large (see
the caption for details). In this case, the simulations confirm
the phenomenon that was pointed out in the previous section,
namely, that the expectation value $\langle I \rangle$ may be
lower than the measured $I({\bf f})$.

It may be seen that for large $N$, all the dots concentrate in
$\langle I \rangle = I({\bf f})$. This is, as pointed out before,
due to the discretization of $\cal D$. If the number of cells $J$
is increased, one needs to go to a larger $N$ to find such a
saturation. On the contrary, for smaller $N$, the simulated
$\langle I \rangle$ lies below its theoretical value. This is a
manifestation of the finite nature of $\epsilon$, and the
phenomenon becomes less evident as $\epsilon$ is lowered.

\section{Discussion}

\label{Discu}

In this work, the probability density $P({\bf q} | {\bf f})$ for
the true distribution ${\bf q}$ given the experimental
frequencies ${\bf f}$ is analyzed. Such a density, it is shown,
may be written as a Gibbs distribution, where the inverse number
of samples plays the role of an effective temperature, and the
Kullback-Leibzig divergence between ${\bf f}$ and ${\bf q}$ is
the equivalent of the energy of state ${\bf q}$. Its study is not
only for academic purposes, but eventually also practical. In the
ideal situation, it would be valuable to calculate $P({\bf q} |
{\bf f})$ while an experiment is being carried out, in order to
know when the number of samples is already enough. The
experimenter may thus decide to give an end to the sampling
process when the width of $P({\bf q} | {\bf f})$ reaches some
acceptable value. For example, someone interested in measuring
the public opinion prior to an election may wonder how many
subjects need to be polled in order to have a reliable estimation
of the forthcoming result. Many times, however, experiments comes
to an end because of other factors (a deadline, or a floor in the
the amount of money, patience or students). An estimation of the
width of $P({\bf q}|{\bf f})$ is valuable even in these cases,
just to provide error bars.

One possibility is to write down the full $P({\bf q} | {\bf f})$.
However, being a function of many variables, this may not be very
practical. A convenient parameter measuring the width of $P({\bf
q } | {\bf f})$ in several directions is the square root of the
corresponding eigenvalues of $\tilde{\Sigma}$. These have been
shown to diminish asymptotically as $1/N$. From the
information-theoretical point of view, a more appealing parameter
is the mean divergence $D$, and its mean quadratic fluctuations.
As is shown in Eq. (\ref{dmed}), for small $N$ such a width
depends on the value of ${\bf f}$. If $N \gg S$, however, both
$\langle D \rangle$ and $\sigma_D$ become independent of ${\bf
f}$ and decrease as $1/N$ (Eq. (\ref{med})). Yet another route is
to work with the function $W({\bf q})$ one is interested in. By
means of Eq. (\ref{gener}), it is possible to decide whether the
term proportional to $1/N$ is only a small correction to $W({\bf
f})$ or, on the contrary, the two terms are comparable. In the
latter case, more measurements should be carried out.

Although some of the expressions presented here are valid for an
arbitrary prior, much of the work deals with the particular case
of Eq. (\ref{prior}). The use of a prior that is essentially a
linear combination of functions of the form (\ref{prior}) has
been proposed \cite{Ilya}, specifically, to be used in the
inference of entropies. For this case, the partition function
should be constructed by applying the same linear superposition to
Eq. (\ref{parti}), and the same holds for Eqs.
(\ref{max}-\ref{completo}).  The calculation of $\langle D
\rangle$ and $\sigma_D$ as derivatives of $F$ is still valid,
whereas Eq. (\ref{result}) should also be averaged.

The analysis of $P({\bf q}|{\bf f})$ carried out in Sect.
\ref{Formu}, and the statistical mechanical description of Sect.
\ref{Tempe} are valid even for small $N$. The fact that $\langle
D \rangle \rightarrow 1/N$ for large $N$ inspires the expansion of
$\langle W \rangle$ of Sect. \ref{Expans}. It should be clear,
nevertheless, that such an expansion is {\bf only} convergent when
$N \gg S$. Actually, Eq. (\ref{result}) is the first order term in
powers of $S/N$, and there is no reason to think that the higher
order terms will be negligible, if such a condition does not
hold. Moreover, it is necessary to have $N f_i \gg 1$ for all
$i$. When $N$ is large enough, one can always define the number
of categories $S$ as to have them all well populated. But for $N
\approx S$ this may well not be the case. The consequences may,
in fact, be quite dramatic. For instance, in the example of the
entropy (Subsect. \ref{ejuno}) one can explicitly see that $f_i$
appears in the denominator  of Eq. (\ref{juli}). In other words,
the result is meaningless if there are empty categories.

However, when the condition $N \gg S$ does hold, Eq.
(\ref{result}) may serve to draw non trivial conclusions. For
instance, it is usually supposed that limited sampling, on
average, flaws the data introducing false correlations. This work
shows this is not necessarily the case: limited sampling may
sometimes, on average, lower the correlations. This is clear in
the simulations of Sect. \ref{Simul}, where finite sampling
results, in mean, in a downwards bias of the mutual information.

\acknowledgments

I thank Ilya Nemenman for his very useful comments and
suggestions. I also thank Alessandro Treves and Stefano Panzeri
for a critical reading of the manuscript. This work has been
carried out with a grant of Fundaci\'on Antorchas.

\appendix

\section{Integrating a power distribution in ${\cal D}$}

Here, Eq. (\ref{integrales}) is derived. An alternative and more
general line of reasoning may be found in \cite{Wolpert}.

The aim is to calculate
\begin{eqnarray}
I_{{\bf m}}^S &=& \int_{{\cal D}} \Pi_{i = 1}^S {\rm d}q_i \
q_i^{m_i}   \\
&=& \int_0^1 {\rm d}q_1 \ q_1^{m_1} \int_0^1 ... {\rm d}q_S \
q_S^{m_S} \delta \left[ \lambda_S \left( 1 - \sum_{j = 1}^S q_j
\right) \right], \nonumber
\end{eqnarray}
where $\lambda_S$ is a constant ensuring that when all $m_i$
vanish, $I_{{\bf 0}}^S$ is the volume of ${\cal D}$. The
supra-index in $I_{{\bf m}}^S$ indicates the dimension of the
vectors ${\bf m}$ and ${\bf q}$.

If $X$ can only take two values, then $S = 2$. In this case,
\cite{gradsh}
\begin{eqnarray}
I_{{\bf m}}^2 &=& \int_0^1 {\rm d}q_1 \ q_1^{m_1} \int_0^1 {\rm
d}q_2 \ q_2^{m_2}  \delta \left[ \lambda_2 \left( 1 - q_1 - q_2
\right) \right], \nonumber \\
&=& \frac {1} {\lambda_2} \int_0^1 {\rm d}q_1 \ q_1^{m_1} (1 -
q_1)^{m_2} \nonumber \\
&=& \frac {1} {\lambda_2} \frac{m_1! m_2!}{(m_1 + m_2 + 1)!}.
\label{cero}
\end{eqnarray}

Now, the hypothesis is made for arbitrary $S$
\begin{equation}
I_{{\bf m}}^S = \frac {1} {\lambda_S} \frac {\Pi_{i = 1}^S m_i!}
{\left(S - 1 + \sum_{j = 1}^S m_j \right)!}. \label{pioui}
\end{equation}
To prove it, one proceeds by complete induction. Eq.
(\ref{pioui}) is assumed true for a given ${\bf m} = (m_1, ...,
m_S)$ and the aim is to prove it for $(m_i, ..., m_{S + 1})$.
Hence
\begin{eqnarray}
I_{(m_1, ..., m_{S + 1})}^{S + 1} &=& \int_{{\cal D}}
\left(\Pi_{i = 1}^{S + 1} {\rm d}q_i \ q_i^{m_i}
\right)  \nonumber \\
&=& \frac{\lambda_{S}}{\lambda_{S + 1}}  I_{(m_1, ..., m_{S -
1})}^{S - 1} \times \nonumber \\
& &  \int_0^{1 - \sum_{i = 1}^S} {\rm d}q_S q_S^{m_S} \left(1 -
\sum_{j = 1}^{S} \right)^{m_{S +
1}} \nonumber \\ & & \ \ \ \ \ \Theta \left(1 - \sum_{j = 1}^{S} \right) \label{uno} \\
&=& \frac{\lambda_{S}}{\lambda_{S+1}} I_{(m_1, ..., m_{S - 1}),
m_S + m_{S + 1} + 1}^S \times \nonumber \\
& & \frac{m_{S}! m_{S + 1}!} {(m_{S} + m_{S + 1} + 1)!} \label{dos}\\
&=& \frac {1}{\lambda_{S+1}} \frac {\Pi_{i = 1}^{S + 1} m_i!}
{\left[(S + 1) - 1 + \sum_{j = 1}^{S + 1} m_j \right]!},
\label{tres}
\end{eqnarray}
where $\Theta(x)$ is Heaviside step function: $\Theta(x) = 1$ if
$x \ge 1$, and $\Theta(x) = 0$ if $x < 0$. When passing from Eq.
(\ref{uno}) to Eq. (\ref{dos}), use was made of the result
(\ref{cero}). Accordingly, (\ref{tres}) derives from the
inductive hypothesis (\ref{pioui}). Since Eq. (\ref{tres})
coincides with (\ref{pioui}) when $S$ is replaced by $S + 1$, the
hypothesis (\ref{pioui}) is proved true.

Finally, to determine $\lambda_S$ one evaluates
\begin{equation}
I_{{\bf 0}}^S = \frac{1}{\lambda_{S} (S - 1)!}.
\end{equation}
The volume of ${\cal D}$ is $\sqrt{S}/(S - 1)!$, as can be
verified, once again, by complete induction. Then $\lambda_S =
1/\sqrt{S}$.


\begin{thebibliography}{999}

\bibitem{stefale} Alessandro Treves, and Stefano Panzeri,
{\it Neural Comp.} {\bf 7} 399 (1995)

\bibitem{network} Stefano Panzeri and Alessandro Treves,
{\it Network} {\bf 7}  87 (1996)

\bibitem{Wolpert} Wolpert David H, and Wolf David R, {\it
Phys. Rev. E} {\bf 52} (6) 6841 (1995)

\bibitem{Kull} Kullback, S., (1968). {\it Information theory and statistics}.
New York: Dover.

\bibitem{Ilya} Ilya Nemenman, Fariel Shafee and William Bialek,
xxx.lanl.gov/abs/physics/0108025 (2001)

\bibitem{Seung} H. S. Seung, H. Sompolinsky and N. Tishby,
{\it Phys. Rev. A} {\bf 45} (8) 6056 (1992)

\bibitem{Bill} William Bialek, Ilya Nemenman and Naftali Tishby,
{\it Neural Comput.}{\bf 13} (11) 2409 (2001)

\bibitem{Kirch} F. Willems, Y. Shtarkov and T. Tjalkens, {\it IEEE Trans. Inf.
Thy.}, {\bf 41}, 653 (1995)

\bibitem{Shur} T. Schurmann and P. Grassberg, {\it Chaos} {\bf 6}
414 (1996)

\bibitem{abra} Abramowitz M and Stegun I. A (Editors), ''Handbook of
mathematical functions'' (Dover, New York, 1972)

\bibitem{gradsh} I. S. Gradshteyn and I. M. Ryzhik, ''Tables of integrals,
series and products'' (Academic Press, San Diego, 1994)

\end{thebibliography}
\end{document}